\documentclass[10pt, letter, twocolumn]{arxiv}

\usepackage{makecell}
\usepackage{kantlipsum, lipsum}
\usepackage{dm-colors}
\usepackage{amsmath}
\usepackage{pstricks, pst-node}
\usepackage{verbatim}
\usepackage{multirow}
\usepackage{scalerel}
\usepackage{booktabs}
\usepackage{enumitem}
\usepackage{xspace}
\usepackage{bm}
\usepackage{bbm}
\usepackage{mathtools}
\usepackage{soul}
\usepackage{epsfig}
\usepackage{graphicx}
\usepackage{csquotes}
\usepackage{setspace}
\usepackage{colortbl}
\usepackage{threeparttable}
\usepackage{tabularx,ragged2e}
\usepackage{placeins}
\usepackage[hang,flushmargin,symbol]{footmisc}

\usepackage[numbers]{natbib}

\usepackage{varioref}
\usepackage[pagebackref=false,breaklinks=false,%
            colorlinks=true,bookmarks=true,citecolor=ourdarkblue,%
            urlcolor=ourdarkblue,linkcolor=ourdarkblue]{hyperref}
\usepackage[noabbrev,capitalize]{cleveref}
\usepackage{etoc}
\usepackage{lineno}

\graphicspath{{figures/}}

\usepackage[misc]{ifsym}

\title{\Large{The Impact of AI Assistance on Radiology Reporting: A Pilot Study Using Simulated AI Draft Reports}}
\author[1\Letter]{Julián N. Acosta}
\author[2]{Siddhant Dogra}
\author[3]{Subathra Adithan}
\author[4]{Kay Wu}
\author[5]{Michael Moritz}
\author[6]{Stephen Kwak}
\author[1]{Pranav Rajpurkar}
\affil[1]{\normalsize Harvard Medical School, Boston, MA, USA \authorcr}
\affil[2]{\normalsize NYU Langone Health, New York, NY, USA \authorcr}
\affil[3]{\normalsize Jawaharlal Institute of Postgraduate Medical Education and Research, Puducherry, India \authorcr}
\affil[4]{\normalsize University of Toronto, Toronto, ON, Canada \authorcr}
\affil[5]{\normalsize Saint Louis University School of Medicine, Saint Louis, MO, USA \authorcr}
\affil[6]{\normalsize Johns Hopkins School of Medicine, Baltimore, MD, USA \authorcr}
\renewcommand{\correspondingauthor}[1]{Corresponding Author: Julián N. Acosta; julian\_acosta@hms.harvard.edu}

\begin{document}

\begin{abstract}
Radiologists face increasing workload pressures amid growing imaging volumes, creating risks of burnout and delayed reporting times. While artificial intelligence (AI) based automated radiology report generation shows promise for reporting workflow optimization, evidence of its real-world impact on clinical accuracy and efficiency remains limited. This study evaluated the effect of draft reports on radiology reporting workflows by conducting a three reader multi-case study comparing standard versus AI-assisted reporting workflows. In both workflows, radiologists reviewed the cases and modified either a standard template (standard workflow) or an AI-generated draft report (AI-assisted workflow) to create the final report. For controlled evaluation, we used GPT-4 to generate simulated AI drafts and deliberately introduced 1-3 errors in half the cases to mimic real AI system performance. The AI-assisted workflow significantly reduced average reporting time from 573 to 435 seconds (p=0.003), without a statistically significant difference in clinically significant errors between workflows. These findings suggest that AI-generated drafts can meaningfully accelerate radiology reporting while maintaining diagnostic accuracy, offering a practical solution to address mounting workload challenges in clinical practice.
\end{abstract}

\maketitle

\section{Introduction}

Radiologists face mounting pressure as imaging volumes surge, requiring them to interpret more studies while maintaining accuracy and speed. This strain has led to concerning burnout rates \citep{Bailey2022-fp} and potential impacts on diagnostic quality. While artificial intelligence (AI) solutions have demonstrated promise in radiology applications like worklist optimization and computer-aided detection, their integration into clinical reporting workflows remains limited.\\
Automated radiology report generation \citep{Liao2023-nr} represents a promising alternative avenue for AI to seamlessly integrate into radiologists' workflows, providing them with AI-generated draft reports that can serve as a starting point in the reporting process. By reducing the time and effort required to generate reports from scratch, AI-assisted reporting has the potential to significantly improve radiologists' efficiency and productivity.\\
Recent studies have focused on evaluating the quality of AI-generated radiology reports using various approaches, such as automated natural language and clinical accuracy metrics \citep{Zhao2023-ag}. Additionally, some studies have conducted manual evaluations to gauge radiologists' perceptions of AI-generated reports \citep{Tu2023-je, Yang2024-uz, Tanno2024-in}. However, the majority of these studies have focused on evaluating the impact of AI assistance on medical image interpretation in experimental settings that differ from radiologists' day-to-day workflows \citep{Yu2024-ts, Bennani2023-oo}. In real-world clinical workflows, image interpretation and report generation often occur in parallel as the radiologist evaluates each case, highlighting the need for studies that closely mimic real-world reporting workflows to fully understand the impact of AI assistance on radiologists' performance and experience.\\ Furthermore, most studies on report generation have focused on simpler modalities like chest X-rays \citep{Tanno2024-in}. However, the potential benefits and challenges of AI-assisted reporting for more complex modalities remain largely unexplored.\\
In this study, we evaluated the impact of AI-generated draft reports on chest CT interpretation using a \textbf{crossover study design} where radiologists modified either standard templates or AI-generated drafts to create final clinically accurate reports, closely mirroring clinical practice. We found that \textbf{AI assistance significantly reduced reporting time while maintaining diagnostic accuracy}, paving the way for larger clinical trials to comprehensively assess the impact of AI-assisted reporting in clinical practice.
\begin{figure}[tp]
\centering
\includegraphics[height=0.8\textheight]{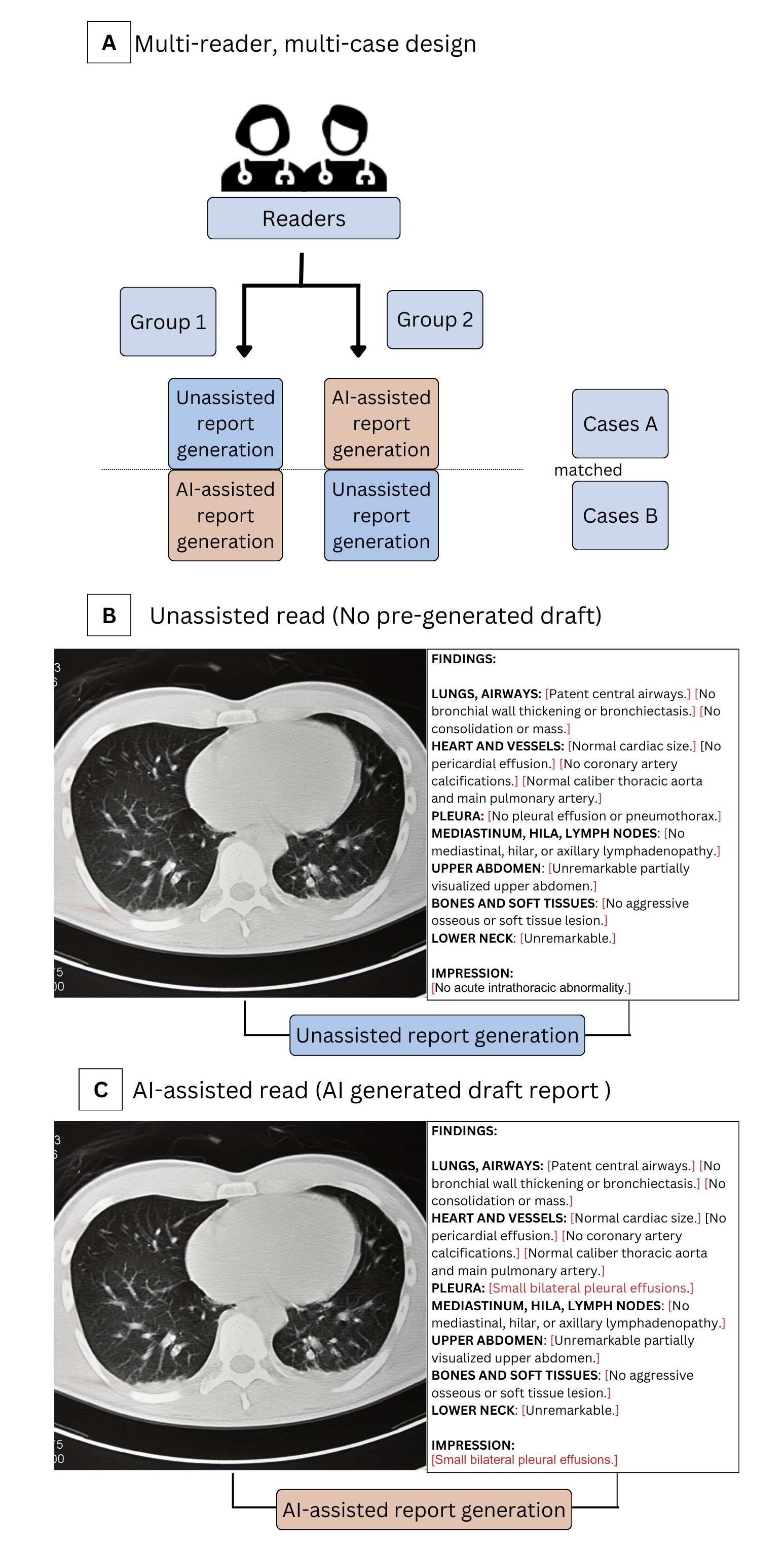} 
\caption{Study overview. (A) Reader assignment and crossover between AI-assisted and unassisted workflows. (B) Unassisted workflow using standard template. (C) AI-assisted workflow with pre-generated drafts.}
\label{fig1}
\end{figure}

\section{Methods}
\begin{figure}[h]
\centering
\includegraphics[width=1.0\columnwidth]{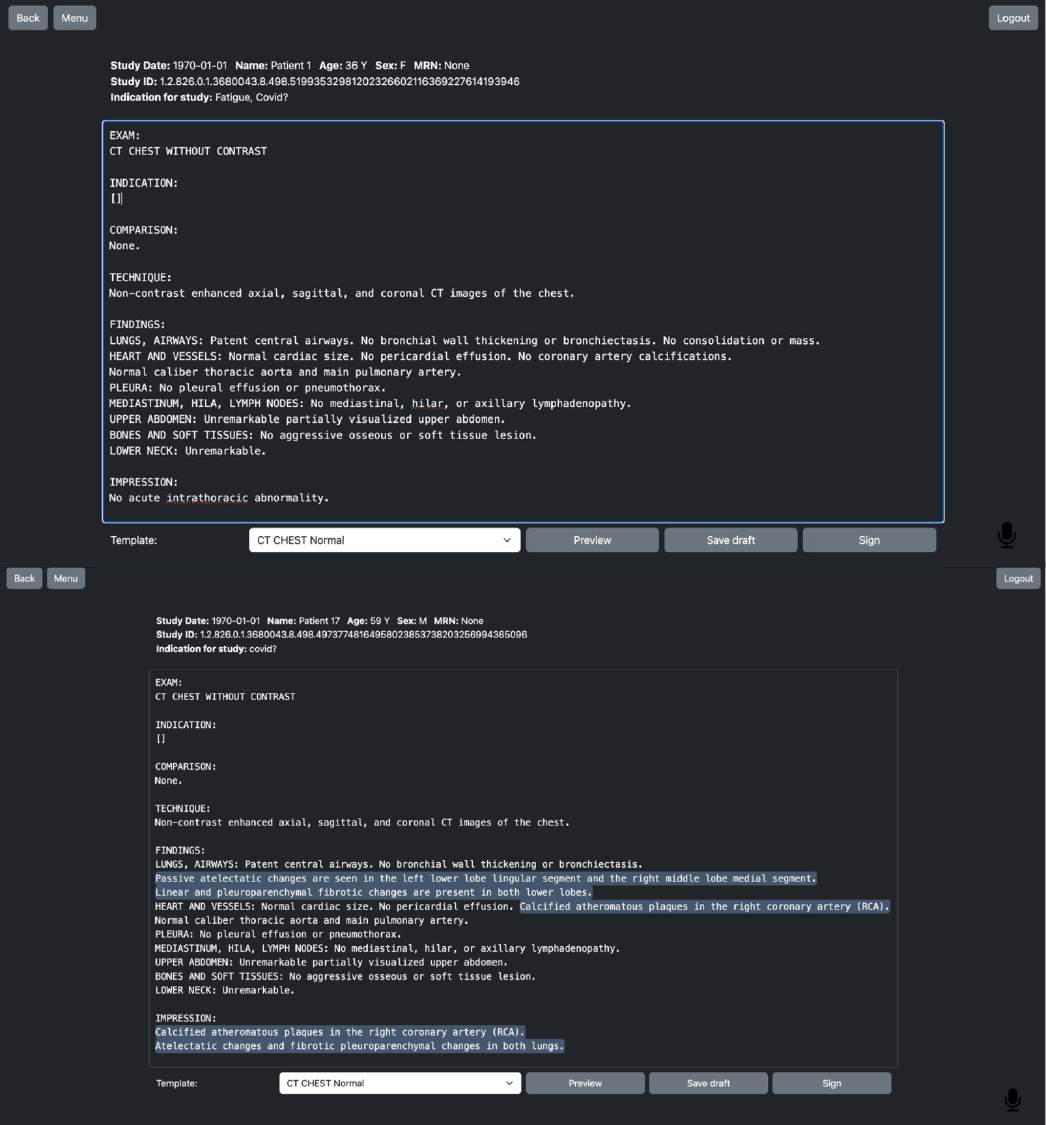}
\caption{Reporting platform. Top: normal negative template case. Bottom: AI-drafted case.}
\label{fig2}
\end{figure}

\textbf{Study Design}
We conducted a three-reader, multi-case crossover study using 20 chest CT scans from the CT-RATE dataset \citep{Hamamci2024-ob}, with each case evaluated under both standard and AI-assisted conditions across different readers.\\\\
\textbf{Cases}
Utilizing the CT-RATE dataset, we randomly selected and curated two groups of 10 chest CT scans each, matched by patient age, sex, and proxies for case complexity, namely the number of findings and the number of impression sentences in the original reports.\\\\
\textbf{Readers}
Three radiology-trained readers participated in the study: one board-certified radiologist and two radiology residents. Each reader evaluated all cases, alternating between standard and AI-assisted workflows to ensure balanced exposure to both conditions.\\\\
\textbf{AI Drafts}
AI drafts were generated by adapting a standard CT chest negative template and incorporating specific findings from the original reports using GPT-4. For half the cases, we deliberately introduced 1-3 errors using GPT-4 to simulate common error patterns observed in AI-generated radiology reports, such as false positives and false negatives \citep{Yu2023-xr}. This simulation approach allowed precise control over error rates, allowing subgroup analysis by number of errors.\\\\
\textbf{Reading Workflow}
Readers evaluated cases in both standard and AI-assisted workflows. In the AI-assisted workflow, readers were provided with pre-generated draft reports where any content indicating abnormal findings was automatically highlighted, regardless of the finding's accuracy, while in the standard workflow, they used normal negative templates (Figure \ref{fig1}), which is standard in radiology practice. For both workflows, readers were instructed to review the imaging findings and modify the provided text as needed to ensure clinical accuracy, following standard radiology practice. The crossover design controlled for potential order effects and individual reader variability while ensuring balanced evaluation across conditions.\\\\
\textbf{Platform Implementation}
We used a custom-built platform intetrating a Python Flask backend with a Postgres SQL database hosted on Google Cloud SQL and a JavaScript/CSS frontend. The platform features a login and signup page, a worklist page listing assigned studies, and a report editing page where participants can view patient data, available clinical information, and a report template or draft report (Figure \ref{fig2}). The platform highlights insertions into the template (positive findings) for easy review and records timestamps at the time of accessing and signing a report. The web incorporates the Open Health Imaging Foundation (OHIF) DICOM viewer, which opens automatically when accessing a case, with images stored and loaded via Google Cloud Health DICOMSTORE.\\\\
\textbf{Endpoints}
We collected two primary outcome measures: reporting time (measured from case opening to report signing) and error assessment (conducted by an independent, experienced radiologist who reviewed all final signed reports for clinically significant errors).\\\\
\textbf{User Experience Assessment}
Following completion of all cases, readers completed a post-experiment survey assessing their experience with AI-assisted reporting. The survey included Likert-scale questions evaluating ease of use, workflow integration, mental effort requirements, and likelihood to recommend the system to colleagues (scored on a 1-10 scale).\\\\
\textbf{Statistical Analysis}
We employed mixed-effects models to analyze both primary outcomes. For error count analysis, we fitted a Poisson mixed-effects model to account for the count nature of the data. Reporting times were analyzed using linear mixed-effects models after log transformation to meet normality assumptions. Both models incorporated fixed effects for patient demographics (age and gender) and case complexity (number of findings), with random intercepts for readers and cases to account for repeated measurements and case-specific variation. Analyses were performed in R (v4.4.2) and the lme4 package.

\section{Results}
\begin{figure}[t]
\centering
\includegraphics[width=0.8\columnwidth]{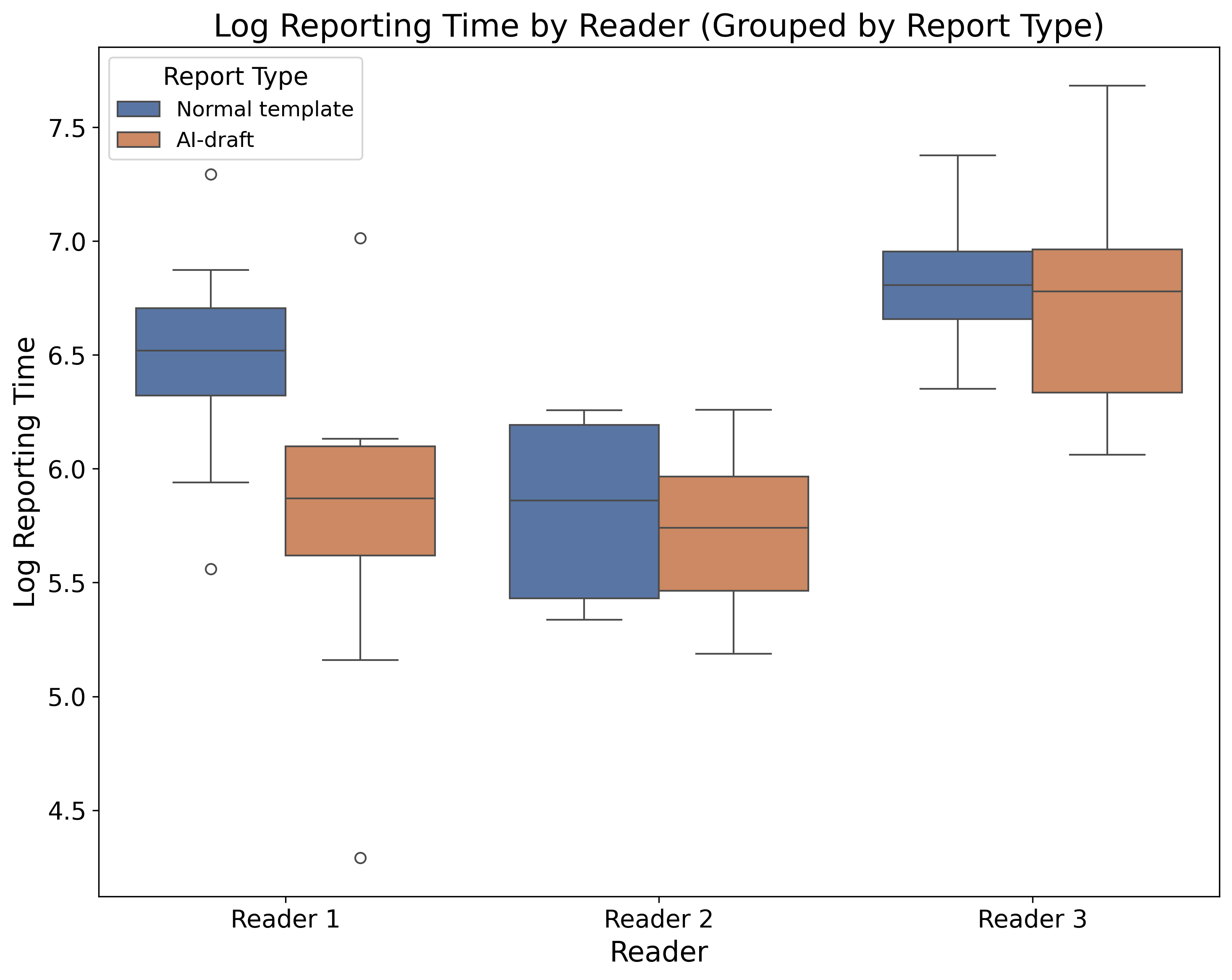}
\caption{Differences in reporting times using AI-drafts by reader.}
\label{fig3}
\end{figure}

\textbf{Study Cases}
The study included 20 cases (50\% female, median age 60.0 years, IQR 33-68) with a final number 59 radiologist-signed reports, with one report excluded due to a data recording error.\\\\ 
\textbf{Clinical Accuracy}
As shown in (Table \ref{table:errors}), the AI-assisted workflow demonstrated a slightly lower mean number of clinically significant errors (0.27±0.52) compared to the standard workflow (0.38±0.78), though this difference did not reach statistical significance in our mixed model analysis.\\\\
\textbf{Reporting Time}
The AI-assisted workflow significantly reduced median reporting time from 573 seconds (IQR 403-895) to 435 seconds (IQR 298-716) (p=0.003), representing a 24\% improvement in efficiency.
Specifically, Reader 1 and Reader 2 demonstrated reduced mean reporting times with AI assistance (717 to 398 seconds and 361 to 322 seconds, respectively), while Reader 3 showed an increase (947 to 1015 seconds). However, all three readers achieved reduced median reporting times (678 to 356 seconds, 354 to 312 seconds, and 904 to 879 seconds). This trend highlights the variability in individual responses but suggests an overall positive impact of AI on workflow efficiency (Figure~\ref{fig3}).\\\\
\textbf{Subgroup Analysis}
In exploratory analyses, we further subdivided the AI-assisted workflow into cases with and without intentionally introduced errors. Subgroup analyses comparing each AI-assisted workflow to standard reporting showed no statistically significant differences, though these analyses were limited by smaller sample sizes in the subgroups.\\\\
\textbf{User Experience}
Post-experiment survey results revealed unanimous positive feedback regarding system usability, with all 3 readers either agreeing or strongly agreeing that the AI-assisted reporting system was easy to use and would be well-integrated into their workflow. Regarding cognitive load, 2 of 3 readers reported that AI-assisted reporting required somewhat less mental effort compared to standard template-based reporting, while 1 of 3 indicated significantly reduced mental effort. However, when asked about likelihood to recommend the system to colleagues, responses showed some variation, with scores of 5, 9, and 10 on a 10-point scale.
\begin{table}[b]
\centering
\caption{Clinically significant errors in preliminary study.}
\begin{tabular}{|p{3cm}|p{2cm}|p{2cm}|} 
\hline
 Clinically significant errors & \textbf{AI-draft (n=30)} & \textbf{Normal template (n=29)} \\ \hline
Mean (SD) & 0.27 (0.52) & 0.38 (0.78) \\ \hline
Median (IQR) & 0 (0-0) & 0 (0-0) \\ \hline
\end{tabular}
\label{table:errors}
\end{table}

\section{Discussion}
\textbf{AI assistance improves reporting efficiency}
Our pilot study suggests that AI-generated draft reports may improve reporting efficiency without compromising diagnostic accuracy, even in the presence of AI errors. The observed 24\% reduction in median reporting time suggests potential for workflow optimization in clinical practice. If replicated in larger studies, this efficiency improvement could substantially reduce radiologists' workload during clinical shifts, allowing more time for complex cases or reducing overall work pressure.\\\\
\textbf{Clinical accuracy remains consistent with AI assistance}
The observation that accuracy remained stable despite the presence of intentional errors in some AI drafts is promising. Our exploratory analyses, though limited by sample size, showed no significant differences in error rates between standard workflow and either type of AI assistance (with or without introduced errors). While encouraging, these findings need validation in larger studies before drawing definitive conclusions about radiologists' ability to maintain vigilance when working with AI-generated content.\\\\
\textbf{Variability across readers}
Individual variability in time savings among our three readers highlights the importance of understanding factors that may influence AI assistance effectiveness, such as experience level, comfort with technology, and personal workflow preferences. Although previous studies have shown remarkable heterogeneity in features influencing AI assistance effects \citep{Yu2024-ts}, larger studies are needed to systematically investigate these individual factors and their impact on AI assistance effectiveness.\\\\
\textbf{User perceptions and adoption considerations}
The unanimous positive feedback regarding system usability and workflow integration from our readers is encouraging, suggesting that AI-generated preliminary reports could be a promising avenue to explore in clinical practice. The consistent reporting of reduced mental effort aligns with our efficiency findings and suggests AI assistance may help address cognitive burden. However, the variability in readers' willingness to recommend a system like this to colleagues reveals an important disconnect between personal usability experiences and broader implementation concerns. This disparity might reflect deeper uncertainties about AI's role in radiology, such as concerns about over-reliance or impact on training, which should be explicitly addressed in future studies.\\\\
\textbf{Limitations}
Our study has several important limitations. With only three readers, our findings may not be generalize to the broader radiologist population. The small sample size limited our statistical power, particularly in the exploratory subgroup analyses. The use of simulated AI drafts rather than actual AI report generation models outputs may not fully reflect real-world performance. Additionally, the artificial setting of a controlled study may not fully reflect real-world clinical practice conditions.\\\\
\textbf{Future Directions}
While our pilot results are encouraging, the next crucial step is conducting a large-scale clinical trial involving multiple readers and cases, using real AI-generated draft reports rather than simulated AI drafts. Such a trial should evaluate not only efficiency and accuracy but also radiologist satisfaction, confidence, and mental effort levels, as well as the impact of different error types and frequencies in AI drafts.\\\\
\textbf{Conclusion}
Our findings suggest that AI assistance in radiology reporting may offer meaningful efficiency gains without compromising diagnostic accuracy, even in the presence of AI errors. However, the observed individual variability and study limitations emphasize the need for larger-scale validation before widespread clinical implementation.

\section*{Disclosures}
JNA, SD, and MM, and PR are part-time employees of a2z Radiology AI. PR is a co-founder of a2z Radiology AI.

\clearpage
\bibliographystyle{sn-mathphys} 
\bibliography{sn-bibliography} 

\end{document}